\documentclass[useAMS,usenatbib]{mn2e}

\usepackage{graphicx}
\usepackage{epstopdf}
\usepackage{amssymb}
\title[An extremely dense group of galaxies at $z=3.09$]{An extremely dense group of massive galaxies at the centre of the protocluster at ${\bf z=3.09}$ in the SSA22 field}
\author[Mariko Kubo et al.]{M. Kubo$^{1, 2}$\thanks{E-mail:
marikubo@icrr.u-tokyo.ac.jp},  
T.~Yamada,$^{2}$ T.~Ichikawa,$^{2}$ M.~Kajisawa,$^{3}$ Y.~Matsuda,$^{4,5}$ 
\newauthor
I.~Tanaka,$^{6}$ H.~Umehata$^{7,8}$\\
$^{1}$Institute for Cosmic Ray Research, University of Tokyo, 5-1-5 Kashiwa-no-Ha, Kashiwa City, Chiba 277-8582, Japan\\
$^{2}$Astronomical Institute, Tohoku University, 6-3 Aoba, Aramaki, Aoba-ku, Sendai, Miyagi 980-8578, Japan \\
$^{3}$Research Centre for Space and Cosmic Evolution, Ehime University, Bunkyo-cho 2-5, Matsuyama 790-8577,  Japan  \\
$^{4}$Chile Observatory, National Astronomical Observatory of Japan, Tokyo 181-8588, Japan \\
$^{5}$SOKENDAI (Graduate University for Advanced Studies), Osawa 2-21-1, Mitaka, Tokyo 181-8588, Japan \\
$^{6}$Subaru Telescope, National Astronomical Observatory of Japan, 650 North A'ohoku Place, Hilo, HI 96720, USA \\
$^{7}$European Southern Observatory, Karl-Schwarzschild-Str. 2, D-85748 Garching, Germany \\
$^{8}$Institute of Astronomy, The University of Tokyo, Mitaka, Tokyo 181-0015, Japan \\
}
\begin{document}

\date{in original form 2014 October 31}

\pagerange{\pageref{firstpage}--\pageref{lastpage}} \pubyear{}

\maketitle

\label{firstpage}

\begin{abstract}
We report the discovery of an extremely dense group of massive galaxies 
at the centre of the protocluster at $z=3.09$ in the SSA22 field 
from near-infrared spectroscopy conducted 
with the Multi-Object InfraRed Camera and Spectrograph (MOIRCS) equipped on the Subaru Telecope.
The newly discovered group comprises seven galaxies confirmed at $z_{\rm spec}\approx3.09$ within 180 kpc 
including five massive objects with the stellar masses larger than $10^{10.5}~M_{\odot}$ 
and is associated with a bright sub-mm source SSA22-AzTEC14. 
The dynamical mass of the group estimated from the line-of-sight 
velocity dispersion of the members is $M_{\rm dyn}\sim1.6\pm0.3\times10^{13}~M_{\odot}$. 
Such a dense group is expected to be very rare at high redshift 
as we found only a few comparable systems in large-volume cosmological simulations. 
Such rare groups in the simulations are hosted in collapsed halos with  
$M_{\rm vir}=10^{13.4}-10^{14.0}~M_{\odot}$ and evolve into the brightest cluster galaxies (BCGs) 
of the most massive clusters at present. 
The observed AzTEC14 group at $z=3.09$ is therefore very likely to be a
proto-BCG in the multiple merger phase. 
The observed total stellar mass of
the group is $5.8^{+5.1}_{-2.0}\times10^{11}~M_{\odot}$.  
It suggests that over half the stellar mass of its descendant had been formed by $z=3$. 
Moreover, we identified over two members for each 
of the four Ly$\alpha$ blobs (LABs) using our new spectroscopic data.
This verifies our previous argument that many of the LABs in the
SSA22 protocluster associated with multiple developed stellar components. 

\end{abstract}


\section{Introduction}

It is still an important open question to understand the early formation history of massive elliptical galaxies. 
Dissipationless or `dry' equal-mass merger is favored to explain dynamically 
hot slow-rotating structure of massive systems while segregation 
of baryon and dark matter in the central part and large phase-space density 
requires dissipational collapse (e.g., \citealt{2006ApJ...636L..81N}). 
Significant size and stellar density evolution from high redshift 
(e.g., \citealt{2005ApJ...626..680D, 2007MNRAS.382..109T, 2008ApJ...677L...5V, 2014ApJ...788...28V})
may be explained by dry minor mergers which `puff-up' 
the galaxy structure to be consistent with those observed at present 
(e.g., \citealt{2009ApJ...697.1290B, 2009ApJ...699L.178N, 2009MNRAS.398..898H}). 
Processes of the formation of massive ellipticals should be also explained 
in the context of the $\Lambda$CDM hierarchical structure growth.

One attractive scenario is the `two-phase' hierarchical multiple mergers 
that was suggested from modern hydrodynamical cosmological 
simulations with very high resolution 
(e.g., \citealt{2003ApJ...590..619M, 2007ApJ...658..710N, 2010ApJ...725.2312O}). 
In these models, in-situ rapid gas accretion and violent star formation 
in the first phase is followed by the secondary phase of multiple 
dry minor or major mergers of satellites which are formed outside 
of the virial radius of a central massive object. 
This scenario explains naturally the rapid and early star formation 
as well as the size and stellar density evolution of massive ellipticals. 
Multiple merger may explain many properties of the most massive galaxies, 
i.e., support by the anisotropic velocity dispersion, smaller ellipticity 
and boxy shape, and relatively diffuse core.

It has become understood that feedback processes, 
especially active galactic nucleus (AGN) feedback, 
play an very important role in the star-formation history
of massive ellipticals (e.g., \citealt{2006MNRAS.365...11C}). 
To avoid the over-cooling problem for massive halos, AGN feedback seems essential. 
It is then a quite important and interesting question to reveal 
how super massive black holes (SMBH) are formed in such hierarchical multiple merging phases. 
If the progenitors of the SMBH of a massive elliptical have already been formed 
in the components of multiple merger, 
the orbital decays of multiple BHs may also affect the final core extension 
of the galaxy (e.g., \citealt{1980Natur.287..307B, 2003ApJ...582..559V}).

There is, on the other hand, still a significant lack of the observational 
constraints for such early history of the massive elliptical galaxy formation. 
In particular, it is not well understood whether such multiple mergers 
really occurred at high redshift, and if occurred, 
what are the properties of the progenitors. 
Herein, we report a discovery or clear identification 
of an extremely dense group of massive galaxies at $z=3$ located 
in the central region of the known most significant protocluster.

In fact, in the recent deep near-infrared (NIR) imaging observations, 
possible hierarchical multiple merger sites have been reported
as the apparent association of massive stellar components in the starburst galaxies at high-redshift.
Double (e.g., \citealt{2006ApJ...640..228T}) or multiple 
(e.g., \citealt{2008ApJ...680..246T, 2012MNRAS.421.2161V})
sources are frequently associated with sub-mm galaxies (SMGs). 
Many of the extended Ly$\alpha$ blobs (LABs) 
also seem to host more than two stellar systems
\citep{2012ApJ...750..116U, 2012ApJ...752...86P, 2013ApJ...771...89O, 2014ApJ...793..114Y}. 
In the local Universe, multiple cores 
which can be the remnants of multiple mergers are sometimes 
found in the brightest cluster galaxies (BCGs) 
(e.g., \citealt{1988ApJ...325...49L, 2002ApJ...577L..89Y, 2003AJ....125..478L, 2003MNRAS.344..110S}).

In order to prove the multiple merging events, however, it is needed 
to spectroscopically identify the components and to examine their dynamical properties. 
Here we report the identification of such dense groups 
by the NIR spectroscopic observations for a sub-mm source, 
SSA22-AzTEC14 \citep{2009Natur.459...61T, 2014MNRAS.440.3462U}, 
as well as the LABs in the core of the SSA22 protocluster at $z=3.09$.  
While the SSA22 protocluster is known as the prominent and the most significant 
density peak of Lyman Break galaxies (LBGs) \citep{1998ApJ...492..428S, 2000ApJ...532..170S}
and Ly$\alpha$ emitters (LAEs) \citep{2000ApJ...532..170S, 2004AJ....128.2073H, 2012AJ....143...79Y},  
significant overdensities of SMGs \citep{2009Natur.459...61T} as well as LABs 
\citep{2004AJ....128..569M, 2011MNRAS.410L..13M} have also been reported. 
In our previous photometric study, it was found that some SMGs 
as well as approximately 40\% of the LABs in this protocluster have over
two stellar counterparts with $K<24$ with photometric redshift 
$z_{\rm phot}\sim 3.1$ \citep{2012ApJ...750..116U}.

We summarise the selection and observations of the targets in Section 2. 
In Section 3, we describe the results of the NIR spectroscopic observations 
and the obtained dynamical properties of the systems.  
Then, we discuss their properties by comparing them 
with similar systems 
in the Millennium simulation \citep{2005Natur.435..629S} in Section 4. 
In Appendix, we give the SED fits of the all galaxies confirmed 
as the members of the galaxy groups studied here. 
In this paper, cosmological parameters 
of $H_0 = 70$ km s$^{-1}$ Mpc$^{-1}$ , $\Omega_{\rm m} =0.3$ 
and $\Omega_{\Lambda} = 0.7$ are assumed.  
The AB magnitude system is used throughout this paper. 

\begin{table*}
 \centering
 \begin{minipage}{160mm}
 \caption{The list of the members of the AzTEC14 group and LABs }
\begin{tabular}{lcccccccc}
\hline \hline
Object & R.A. & Dec  & $K_{\rm tot}$ & $J-K$ & $M_{\ast}$ & $z_{\rm spec}$   & Exptime\footnote{Exposure times of our NIR spectroscopic observations}  & Ref
\footnote{The reference of the redshifts are \citet{2003ApJ...592..728S} (S03), \citet{2010MNRAS.402.2245W} (W10),
\citet{2013ApJ...767...48M} (M13), \citet{2014ApJ...795...33E} (E14)  and \citet{2015ApJ...799...38K} (K15). 
The redshifts cited from S03 and W10 were obtained with the optical spectroscopy 
while others were obtained with the NIR spectroscopy. }
  \\ 
 & (J2000.0) & (J2000.0) & (mag) & (mag) & ($10^{10}~M_{\odot}$)  &   & (ks) & \\
\hline
\multicolumn{9}{c}{AzTEC 14 group} \\
 \hline
Az14-K15a & 22:17:37.3 & +00:18:23.2 & 22.5 & 2.8 &$ 8.0_{- 2.9 }^{+ 5.7 } $ & 3.0851 $\pm$ 0.0001 & 13.6 & K15 \\
                            &                      &                        &          & &        &  3.0926 $\pm$ 0.0003 & & \\
Az14-K15b   & 22:17:36.8 & +00:18:18.2 & 23.1 & 1.6 & $ 9.3_{- 5.0 }^{+ 19.7 } $ &  3.0854 $\pm$ 0.0003 & 13.6 & K15 \\
Az14-K15c   & 22:17:37.3 & +00:18:16.0 & 21.6 & 2.8 & $ 25.4_{- 5.8 }^{+ 7.3} $ &  2.7, $3.0-3.15$\footnote{ There is a large uncertainty since the redshift of this object was measured with the Balmer/4000 \AA~breaks of its stacked continuum spectrum. }
        &  13.0 & K15 \\
Az14-K15d   & 22:17:37.1 & +00:18:17.9 & 23.3 & 1.1 & $ 5.3_{- 2.9 }^{+ 5.7 } $ &   3.0774 $\pm$ 0.0003   & 20.0 & this study\\
Az14-K15e   & 22:17:37.1 & +00:18:22.4 & 23.4 & 2.0 & $ 7.2_{- 5.2 }^{+9.8} $ &  3.0925 $\pm$ 0.0002     &  20.0 & this study \\
Az14-K15f   & 22:17:36.9 & +00:18:38.0 & 23.1 & 2.3 & $ 1.7_{- 0.8 }^{+ 2.2 } $ &  3.0866 $\pm$ 0.0002     &  20.0 & this study \\
MD048  & 22:17:37.7 & +00:18:20.6 &  24.1& ... & $ 1.1_{- 0.7 }^{+ 1.1 } $ &  3.086 & ... & S03 \\
\hline
\multicolumn{9}{c}{LAB01} \\
\hline
LAB01-K15a & 22:17:26.1 & +00:12:32.3 & 22.1 & 1.3 & $ 9.5_{- 2.5 }^{+ 7.5 } $  & 3.1000 $\pm$ 0.0003 &  14.0 & K15 \\
LAB01-K15b & 22:17:25.7 & +00:12:38.7 & 23.6 & 0.3 &$ 1.1_{- 0.5 }^{+ 0.6 } $ &  3.1007 $\pm$ 0.0002; &  14.0 & K15 \\
LAB01-K15c & 22:17:26.0 & +00:12:36.4 & 22.9 & 1.4 &$ 8.1_{- 2.4 }^{+ 2.9} $ & 3.099 & 7.6 & W10 \\
C11 & 22:17:25.7 & +00:12:34.7 & 23.2 & 1.3 & $ 1.1_{- 0.3 }^{+ 2.1} $ &  3.0999 $\pm$ 0.0003 & ...  & M13 \\
C15 & 22:17:26.2 & +00:12:54.7 & ... & ... & ... &  3.0986 $\pm$ 0.0003 & ...  & M13 \\
\hline
\multicolumn{9}{c}{LAB12} \\
\hline
LAB12-K15 & 22:17:32.0 & +00:16:55.5 & 22.2 & 2.6 & $ 11.5_{- 4.9 }^{+ 23.9 } $ &  3.0909$\pm$ 0.0004 & 13.6 & K15 \\
SSA22-012 & 22:17:31.714 & +00:16:57.262 & 23.3 & 1.2 & $2.0_{- 1.1 }^{+ 2.3 } $ &  3.0902 &  ... & E14\\
\hline
\multicolumn{9}{c}{LAB16} \\
\hline
LAB16-K15 & 22:17:24.85 & +00:11:17.60 & 23.1 & 0.8 &  $ 4.1_{- 2.0 }^{+ 3.0 } $ & 3.0689$\pm$ 0.0002  & 14.0  & K15 \\
SSA22-006 & 22:17:24.833 & +00:11:16.002 & $>$25.2 &  ... & ... &  3.0691 & ...  & E14 \\
\hline
\multicolumn{9}{c}{LAB30} \\
\hline
LAB30-K15a & 22:17:32.45 & +00:11:32.87 & 23.0 & 0.9 & $4.2_{-0.6}^{+0.2} $ &   3.0687$\pm$ 0.0004 &  14.0  & K15 \\
LAB30-K15b & 22:17:32.52 & +00:11:31.22  & 23.4 & 0.4 & $4.4_{-0.3}^{+0.2} $ &  3.0680$\pm$ 0.0003 &  14.0  & K15 \\
SSA22-001\footnote{This object is likely to be the same as LAB30-K15a} & 22:17:32.453 & +00:11:33.513 & ... & ... & ... &  3.0690 &... & E14 \\
\hline
\end{tabular}
\end{minipage}
\label{tab:tableobjects2}
\end{table*}

\section[Selection and observations of multiple galaxies]{Selection and observations of the multiple merger candidates}
The targets are the counterparts of   
a SMG and LABs in the SSA22 protocluster at $z=3.09$, 
which were selected in our own deep $K_s$-band images 
obtained with the multi-object infrared camera and spectrograph (MOIRCS) 
equipped on the Subaru Telescope.
The details of the dataset and reduction procedures  
are given in \citet{2012ApJ...750..116U} and \citet{2013ApJ...778..170K}. 
Briefly, we selected sources with $K <24$ ($5\sigma$ detection limit) 
and $2.6<z_{\rm phot}<3.6$ as candidate protocluster galaxies.   
In addition, distant red galaxies (DRGs; $J-K>1.4$, \citealt{2003ApJ...587L..79F})
were selected as the candidate protocluster galaxies. 
The photometric redshifts of the galaxies were estimated 
from spectral energy distribution (SED) fitting 
of their fluxes at $u^{\star}BVRi'z'JHK$, 3.6, 4.5, 5.8 and 8.0 $\mu$m-bands 
(taken by \citealt{2004AJ....128.2073H, 2004AJ....128..569M, 2009ApJ...692.1561W, 2012ApJ...750..116U}) 
with the stellar population synthesis models of \citet{2003MNRAS.344.1000B} 
through a standard $\chi^2$ minimisation procedure. 
We adopt the Chabrier Initial Mass Function (IMF) \citep{2003PASP..115..763C} 
in this paper and \citet{2015ApJ...799...38K} (here after K15) 
while the Salpeter IMF \citep{1955ApJ...121..161S} 
was used in \citet{2012ApJ...750..116U} and \citet{2013ApJ...778..170K}. 
Adopting the Chabrier IMF, our sample selection 
is nearly complete for the galaxies 
with $M_{\ast}\geq1\times10^{10}~ M_{\odot}$ at $z\sim3$ 
if the sample is dominated by normal star forming galaxies. 
However, as we describe in \S 3, our sample is biased 
to rare dusty and/or passively evolving galaxies 
which have larger stellar mass to light ratios than 
those of normal star forming galaxies. 
Therefore we take the conservative stellar mass limit 
based on the observed stellar populations 
of the galaxies in each galaxy group in later.

It was reported in \citet{2012ApJ...750..116U} that  
40\% of the LABs in the SSA22 protocluster  
have over two $K$-band counterparts with $2.6<z_{\rm phot}<3.6$ 
within their Ly$\alpha$ nebulae. 
They also reported the overdensities of hyper extremely red objects 
(HEROs; $J-K>2.1$, \citealt{2001ApJ...558L..87T}) 
around the AzTEC/ASTE 1.1 mm sources in the SSA22 field 
\citep{2009Natur.459...61T}. 
The most extreme cases among them are LAB01, LAB02, the largest LABs,  
and SSA22-AzTEC14 \citep{2014MNRAS.440.3462U} 
which are plausibly associated with more than five counterparts 
within the spatial extents of $\approx$150 kpc 
(\citealt{2012ApJ...750..116U}, note that the source ID of AzTEC14 
was redefined from AzTEC20). 

\begin{figure*}
\includegraphics[width=170mm]{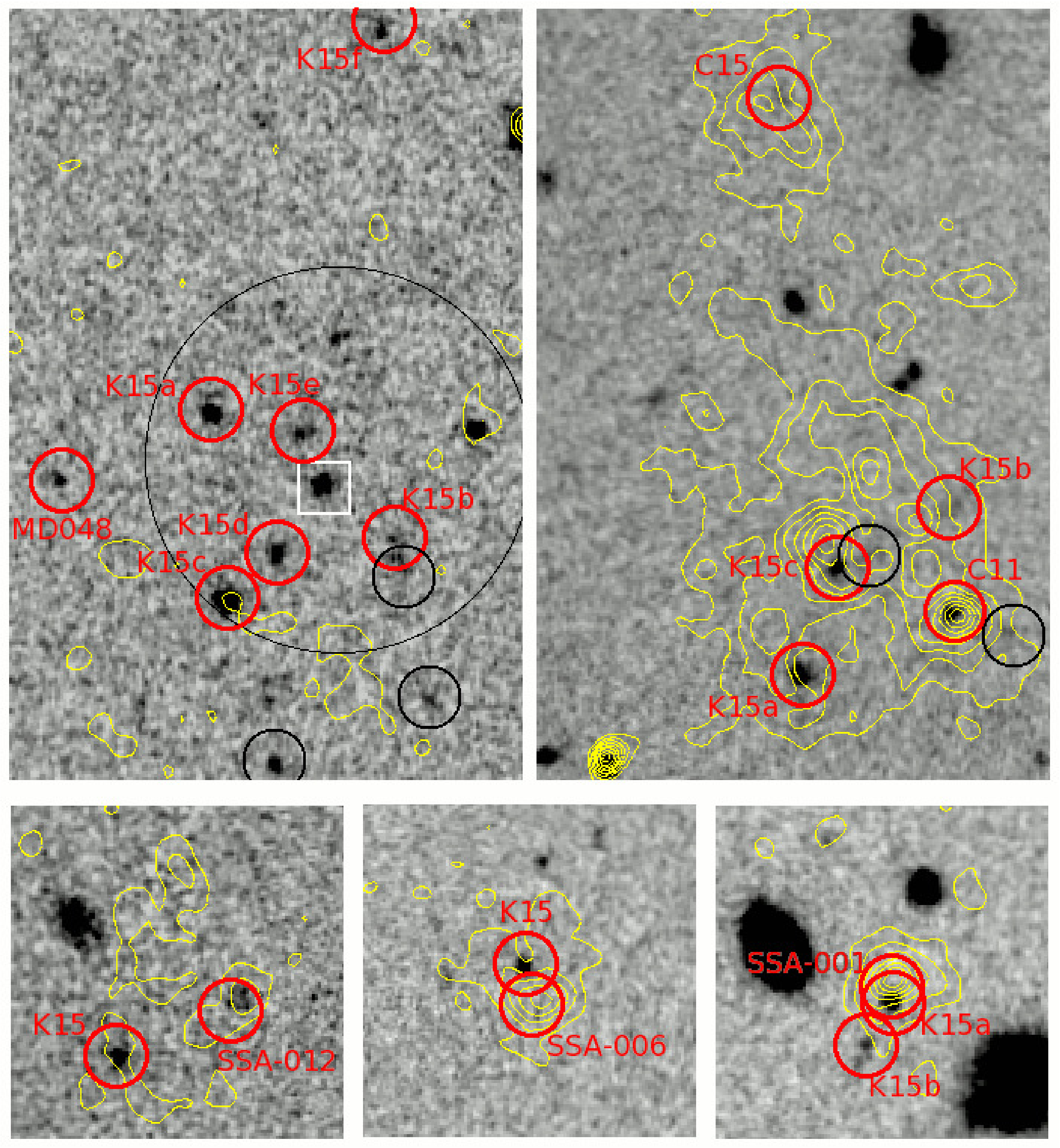}
\caption{The $K_s$-band images of the AzTEC14 group ({\it top left}), LAB01({\it top right}), 
LAB12 ({\it bottom left}),  LAB16 ({\it bottom centre})  and LAB30 ({\it bottom right}). 
The sizes of the images are $20.0\times30.0$ arcsec 
($\approx150$ kpc $\times$ 230 kpc in physical) for the AzTEC14 group and LAB01, 
and $13.0\times13.0$ arcsec for LAB12, LAB16  and  LAB30. 
The thick red circles with IDs show the objects with $z_{\rm spec}=3.07-3.10$. 
Their IDs correspond to those in Table \ref{tab:tableobjects2}. 
The black circles indicate the galaxies with $2.6<z_{\rm phot}<3.6$.  
The white square shows the foreground object with $z_{\rm spec} =0.5763$ (K15). 
The yellow contours show Ly$\alpha$ isophotal area obtained with the 
$BV$-band and $NB497$-band images taken with the Subaru Suprime-Cam \citep{2004AJ....128..569M} 
while the coordinates on the $K_s$-band images were calibrated to those on the Suprime-Cam images. 
The black large circle on the AzTEC14 group shows the beam-size
of the ASTE/AzTEC 1.1 mm source \citep{2014MNRAS.440.3462U}.
 } 
\label{fig:mmgimages}
\end{figure*}

We then conducted NIR spectroscopic observations 
to confirm the redshifts of the candidate protocluster galaxies.  
Since Ly$\alpha$ is a resonance line, other nebulae emission lines are more suitable 
for obtaining the systemic redshifts of the stellar components associated with LABs. 
For galaxies at $z\approx 3.09$,  [O{\footnotesize III}] $\lambda5007$ emission line
is the strongest observable emission line in the NIR. 
The observations were conducted with MOIRCS  
in September and October 2012 (K15) 
and in June and July 2014. 
We observed the counterparts of SSA22-AzTEC14 and LAB01 intensively. 
In addition to these unique objects, LAB12, LAB16 and LAB30 were observed. 
The details of the observations and data reduction procedures,  
and the complete list of our NIR spectroscopic sample in the 2012 run are given in K15. 
To summarise, the detection limit of the data taken in the 2012 run was $\sim1-2\times10^{-17}$ 
erg s$^{-1}$ cm$^{-2}$ ($\sim 3\sigma$ of the background at $\sim$2.05 $\mu$m) 
when the exposure time was $13.0-14.0$ ks for each mask. 
During the observation in June 2014, the full width at half maximum (FWHM) 
of the point spread function (PSF) ranged from 0.35 to 0.4 arcsec. 
The total exposure time was 20.0 ks for this mask. 
Consequently, emission lines as faint as $6.5\times10^{-18}$ erg s$^{-1}$ cm$^{-2}$ were detected. 
On the other hand, in July 2014, humidity was high  
and the FWHM of the PSF during the observation ranged from 0.5 to 0.8 arcsec. 
Owing to the poor conditions and short exposure time of 7.6 ks, 
the limiting flux value of the taken in that run was 
 $\sim2-3\times10^{-17}$ erg s$^{-1}$ cm$^{-2}$,  
which was too shallow to detect the redshifted 
[O{\footnotesize III}] $\lambda5007$ emission lines of the galaxies at $z\sim 3$ at most. 

\begin{figure*}
\begin{center}
\includegraphics[width=178mm]{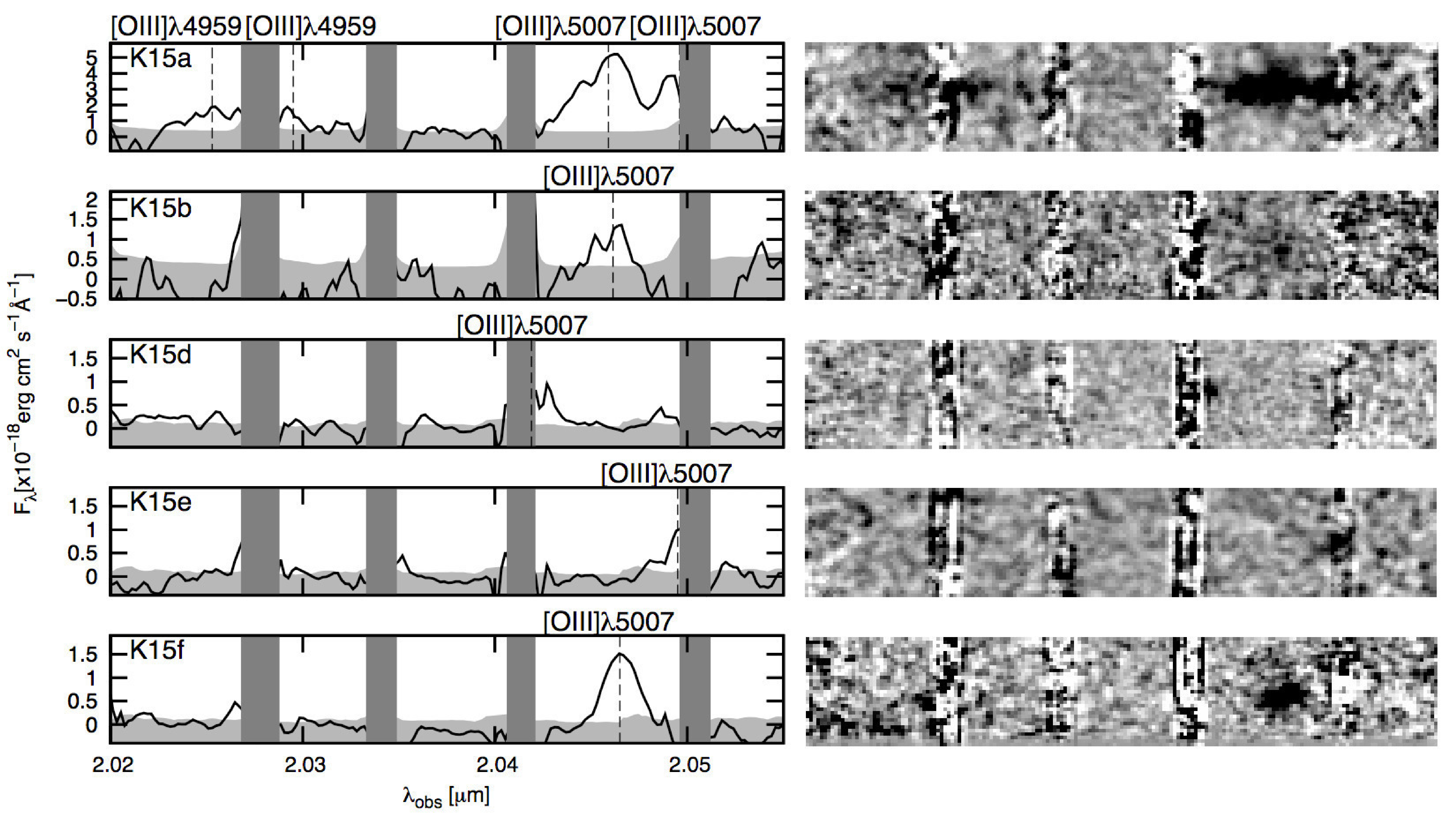}
\caption{
Left panels: The solid lines show the one-dimensional spectra 
of Az14-K15a, b, d, e and f at 2.020 to 2.055 $\mu$m from top to bottom.
The dashed vertical lines indicate the centres of the detected emission lines. 
Only the [O{\scriptsize III}] $\lambda5007$ emission lines are detected at most while 
the [O{\scriptsize III}] $\lambda4959$ are also detected for Az14-K15a. 
The grey shaded regions show the 1$\sigma$ background noise at each point. 
Dark grey areas mask the regions buried in strong OH sky emission lines. 
Right panels: The images of the spectra correspond to the left panels. 
} 
\label{fig:specimage}
\end{center} 
\end{figure*}

\begin{figure*}
\begin{center}
\includegraphics[width=140mm]{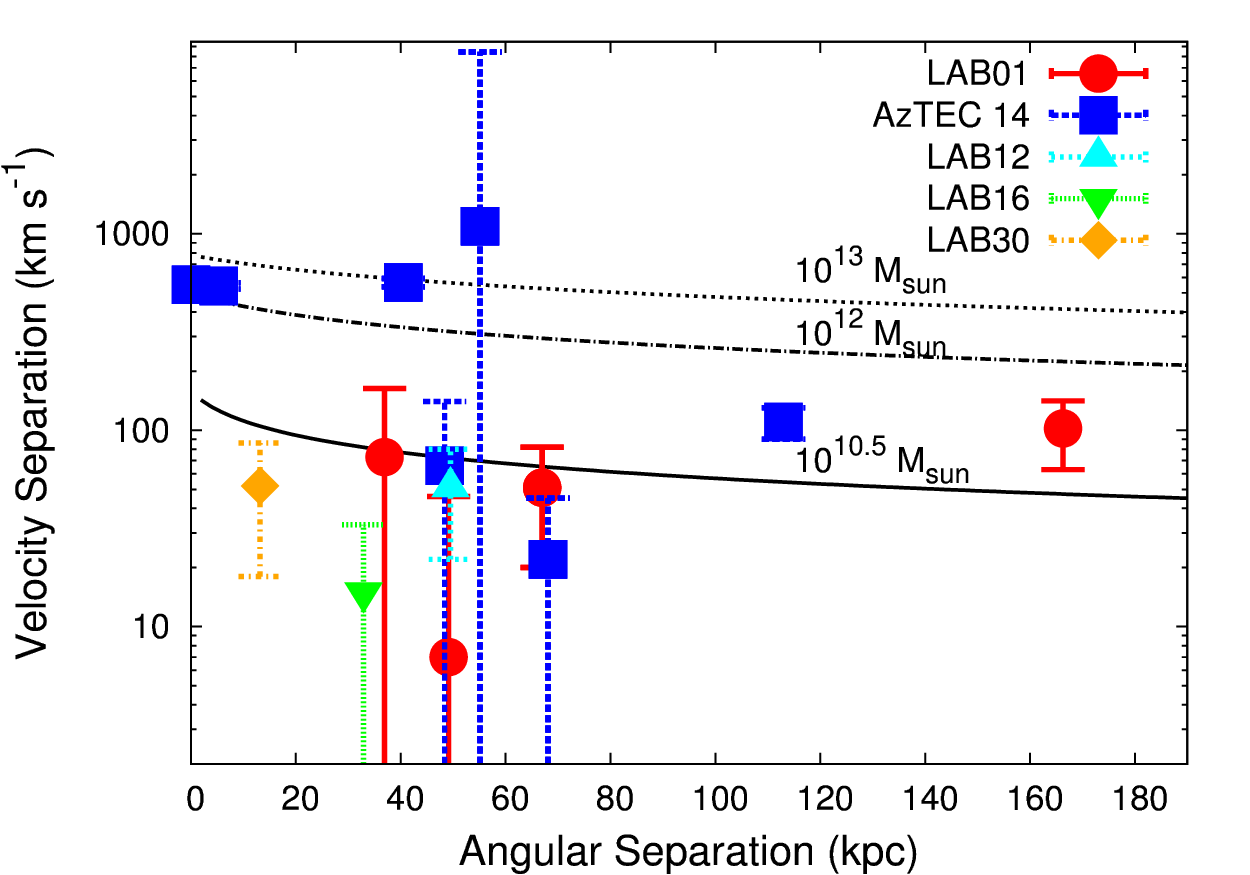}
\caption{
The line-of-sight velocity and spatial distributions of the group members. 
Blue squares, red circles, a cyan triangle, a green reversed triangle and an orange diamond 
show angular spatial separations and line-of-sight velocity offsets of the counterparts 
from the central counterparts of the AzTEC14 group, LAB01, LAB12,  LAB16 and LAB30, respectively. 
The solid, dashed dotted and dotted curves indicate the escape velocities 
from the objects with $M_{\rm halo}=10^{10.5}$, $10^{12}$ and $10^{13}~M_{\odot}$, respectively, 
calculated assuming the Navarro-Frenk-White (NFW; \citealt{1997ApJ...490..493N}) mass profile. 
We assume the concentration parameter $c=5$ for the halo with $M_{\rm halo}=10^{13}~M_{\odot}$, 
and $c=4.5$ for the halos with $M_{\rm halo}=10^{10.5}$ and $10^{12} ~M_{\odot}$
predicted in \citet{2011ApJ...740..102K}. 
} 
\label{fig:space-velocity}
\end{center} 
\end{figure*}

\begin{table}
 \begin{minipage}{83mm}
 \centering
 \caption{The list of the newly confirmed counterparts of the AzTEC14 group}
\begin{tabular}{lcc}
\hline \hline
Object  & $z_{\rm [OIII]}$ & $F_{[\rm OIII] \lambda 5007}$ \\
           &    & ($10^{-17}$ erg/s/cm$^2$) \\
\hline
Az14-K15d & 3.0774 $\pm$ 0.0003  & $2.2\pm0.2$ \\
Az14-K15e & 3.0925 $\pm$ 0.0002  & $1.5\pm0.1$ \\
Az14-K15f & 3.0866 $\pm$ 0.0002  & $3.2\pm0.4$ \\
\hline
\end{tabular}
\label{tab:tableobjects1}
 \end{minipage}
\end{table}

\section{Results}
\subsection{Spectroscopic redshifts}

We summarise the spectroscopic redshifts of the SMG and LABs 
studied here in Table \ref{tab:tableobjects2}. 
The redshifts and flux values of the [O{\footnotesize III}] $\lambda$5007 emission lines 
of the newly confirmed counterparts are listed in Table \ref{tab:tableobjects1}.  
We measure the redshifts and flux values 
by fitting the emission lines with the Gaussian functions. 
Although only one emission line was detected for most of the targets,
the photometric redshifts suggest that they are [O{\footnotesize III] $\lambda$5007 at $z\sim3.09$.
Fig. \ref{fig:specimage} shows the spectra of the counterparts of SSA22-AzTEC14.  
About half the emission lines of Az14-K15d and -K15e 
are likely hidden by OH sky emission lines. 
Their redshifts are estimated by fitting their line profiles 
taking the hidden regions into account.  
Besides our own redshifts, we cite the spectroscopic redshifts from  
\citet{2003ApJ...592..728S}, \citet{2010MNRAS.402.2245W}, 
\citet{2013ApJ...767...48M} and \citet{2014ApJ...795...33E}.
 Here we treat the spectroscopically confirmed counterparts as the group members.  
Two or more galaxies are confirmed for the counterparts of SSA22-AzTEC14 
(hereafter referred to the AzTEC14 group), LAB01, LAB12, LAB16 and LAB30.  

The stellar mass values of the galaxies are estimated from the SED fitting 
with the same procedure as that in K15.
Before the fitting, we subtracted the [O{\footnotesize III}]
$\lambda$5007 emission-line contribution from the $K$-band magnitude,
which results in a correction of $< 0.1$ mag.
The observed and best-fit SEDs of all the counterparts are shown in the Appendix. 
Az14-K15a and LAB12-K15 are the AGNs detected in X-ray \citep{2009MNRAS.400..299L}
but they have no significant AGN feature in their SEDs
and are well fitted with the SED models of galaxies. 
The errors of the stellar masses show a 68\% confidence level in the probability distribution 
for the stellar masses calculated from a minimum $\chi^2$ value for each object. 
The uncertainties in the redshifts have only negligible effects 
on those in the SED fits and stellar masses.
Assuming that the typical stellar population of the galaxies 
in each group is given by those of the confirmed counterparts, 
the completeness may decline substantially 
for galaxies with $M_{\ast}\leq4\times10^{10}~M_{\odot}$
and $M_{\ast}\leq2-3\times10^{10}~M_{\odot}$ 
in the AzTEC14 group and the LABs, respectively. 

Fig. \ref{fig:mmgimages} shows the MOIRCS $K_s$-band images
of the AzTEC14 group and the LABs. 
The yellow contours show the isophotal areas 
of Ly$\alpha$ emission at $z=3.09$, which was 
drawn from the $NB497$-band images subtracted with the $BV$-band images
both obtained using the Subaru Suprime-Cam \citep{2004AJ....128..569M}. 
The objects whose redshifts were confirmed in other studies  
sometimes show no significant $K_s$-band counterpart 
while they are identified as  
i.e., LBGs or the surface brightness peaks of Ly$\alpha$ nebulae.  
Surface number densities of the galaxies with $K<24$ 
at $z_{\rm spec}\approx3.09$ are 232 arcmin$^{-2}$ for the AzTEC14 group 
excluding Az14-K15f (60 arcmin$^{-2}$, if we include Az14-K15f) 
and 132 arcmin$^{-2}$ for LAB01, 
estimated by taking the aperture areas containing all the group 
members with $K<24$. 
These surface number densities are 60 and 30 times 
larger than those in the central part of the SSA22 protocluster studied with MOIRCS,  
which is 1.7 times larger than that in the blank field \citep{2013ApJ...778..170K}. 
Such a dense group of physically associated massive galaxies has never been confirmed at $z>3$.  
Similarly \citet{2014MNRAS.445..201M} found 
a plausible lensed compact group of galaxies at $z\sim 2.9$ 
but based on a photometric lensing model. 

The following describes the properties of the individual objects studied here.
In this paper, we do not refer the total star formation rates (SFRs) of the groups 
estimated from the SED fitting since there are 
large uncertainties due to the faint counterparts. 

\subsubsection{The AzTEC14 group} 

In addition to K15, we newly confirmed three counterparts.  
The redshifts and flux values of the [O{\footnotesize III}] $\lambda$5007 emission lines 
of the newly confirmed counterparts are listed in Table \ref{tab:tableobjects1}.  
In total, seven galaxies at $z_{\rm spec}\approx3.09$ were confirmed. 
Note that there is a large redshift uncertainty for Az14-K15c 
since its redshift was measured with the Balmer/4000 \AA~breaks 
of its stacked continuum spectrum  
but here we treat this object as a very plausible member of the group. 
Fig. \ref{fig:specimage} shows the spectra of the members of the AzTEC14 group.  
Az14-K15a shows the extended emission lines with double peaks 
which can be originated in a dual AGN and/or the gas motion
in the narrow line region of an AGN;  
this object is an AGN detected in X-ray \citep{2009MNRAS.400..299L}. 

The members of the AzTEC14 group are dominated by extremely red objects 
classified as DRGs and HEROs \citep{2012ApJ...750..116U}. 
Some of them are dusty red galaxies since 
they are plausible counterparts of an 1.1 mm source 
and an IRAM PdBI source is certainly identified 
at the location of Az14-K15e (Umehata et al. in prep).   
On the other hand, there are also some old red galaxies 
like Az14-K15a and Az14-K15c 
which are well characterized as massive quiescent galaxies in K15. 
The total SFR estimated from the flux values 
of the [O{\footnotesize III}]  $\lambda$5007 emission lines 
is $85.1\pm8.7$ $M_{\odot}$ yr$^{-1}$, 
adopting $ {\rm SFR_{\rm H \alpha}}(M_{\odot} $yr$^{-1}) =L({\rm H}\alpha) /1.12\times 10^{41}~{\rm ergs~ s}^{-1}$
\citep{1983ApJ...272...54K} and assuming that the [O{\footnotesize III}] $\lambda$5007 to H$\alpha$ ratio is unity, 
where the no reddened H$\alpha$/H$\beta$ ratio is 2.88 \citep{1989agna.book.....O} 
and [O{\footnotesize III}]/H$\beta$ ratios are empirically $\sim0.3-30$ for $z=2-3$ 
star forming galaxies (e.g., \citealt{2014ApJ...795..165S}). 
Here, the extinction of and the contribution of AGNs 
to the [O{\footnotesize III}] $\lambda$5007 emission lines are ignored. 

\subsubsection{LAB01}

In total, four galaxies were confirmed as the counterparts of LAB01. 
The redshifts of LAB01-K15a, b and C11 were determined
from the [O{\footnotesize III}] $\lambda$5007 emission lines. 
For LAB01-K15c, we use $z_G =3.099$ obtained by \citet{2010MNRAS.402.2245W} 
as the systemic redshift,
which was measured by fitting the Ly$\alpha$ line profile with Gaussian plus absorption features
of the spectrum of the optical source detected at the position of LAB01-K15c
(R3 in \citealt{2010MNRAS.402.2245W}). 
We conducted the NIR spectroscopy of this object in July 2014 but failed to detect  
any emission line probably due to the poor condition. 
C15 within the north nebula of LAB01 was also confirmed 
at the redshift close to those of the counterparts of LAB01. 
The redshifts were also measured with Ly$\alpha$ for LAB01-C11 and C15,
which offset from the systemic redshifts measured with [O{\footnotesize III}] $\lambda$5007
by $-51.3\pm42.1$ and $+5.8\pm32.9$ km s$^{-1}$, respectively \citep{2013ApJ...767...48M}. 
These offsets are lower than those of typical LAEs ($\sim200$ km s$^{-1}$ 
\citealt{2014ApJ...795...33E}).

SFR$_{[{\rm O III}]}$ of LAB01 is $64.6\pm6.5~M_{\odot}$ yr$^{-1}$ in total.
Note that there may be dust obscured star formation activities and/or QSOs since 
LAB01-K15a and LAB01-K15c are detected 
in {\it Spitzer} MIPS $24~\mu$m \citep{2009ApJ...692.1561W} 
and JCMT/SCUBA-2 850 $\mu$m \citep{2014ApJ...793...22G}.  

\subsubsection{LAB12, LAB16 and LAB30}

The redshifts of the counterparts for LAB12,
LAB16 and LAB30 in Table 1 were determined 
from the [O{\footnotesize III}] $\lambda$5007 emission lines.
The redshifts measured with Ly$\alpha$ offset 
from the systemic redshifts measured with [O{\footnotesize III}] $\lambda$5007
by $+278\pm50$ km s$^{-1}$, $+508\pm68 $ km s$^{-1}$ and $+442\pm19$ km s$^{-1}$ 
for LAB12-S012, LAB16-S006 and LAB30-S001, respectively \citep{2014ApJ...795...33E}, 
which are similar to those of the typical LBGs. 

The total SFR$_{[{\rm O III}]}$ of LAB12, LAB16 and LAB30 are 
$9.4\pm1.9$, $18.1\pm1.2$ and $5.8\pm1.4~M_{\odot}$ yr$^{-1}$, respectively. 
LAB12-K15 is characterized as a massive quiescent galaxy with an AGN (K15). 

\subsection{Velocity distributions on the groups}

Fig. \ref{fig:space-velocity} shows the angular spatial separations 
and line-of-sight velocity offsets of the group members from their centres,  
where the centre of a group is defined as the location and redshift of the group member 
with the largest stellar mass. 
We treat Az14-K15a as the centre of the AzTEC14 group 
since the redshift uncertainty of Az14-K15c is large. 
Nonetheless, its redshift is close to the median redshift of the AzTEC14 group. 

The solid, dashed dotted and dotted curves in Fig. \ref{fig:space-velocity} 
indicate the escape velocities from the halos with halo mass $M_{\rm h}=10^{10.5}$,  
$10^{12}$ and $10^{13}~M_{\odot}$, respectively, 
calculated assuming the Navarro-Frenk-White (NFW; \citealt{1997ApJ...490..493N}) mass profile
with the concentration parameter $c=5$ for $M_{\rm h}=10^{13}~M_{\odot}$,
and $c=4.5$ for $M_{\rm h}=10^{10.5}~M_{\odot}$ and $10^{12} ~M_{\odot}$. 
We assume the concentration parameters at $z\sim3$ predicted in \citet{2011ApJ...740..102K}. 
Note that here we ignore the line-of-sight separations 
and the velocities in the transverse direction. 
The objects below these curves can be dynamically bound
to one system with these mass values. 
The AzTEC14 group can be bound to an object with mass larger than $10^{13}~M_{\odot}$.  
Assuming that the AzTEC14 group is spherical symmetry and in the dynamical equilibrium,
its dynamical mass
is estimated to be $M_{\rm dyn}=3\sigma_v^2 R /G\sim1.6\pm0.3\times10^{13}~M_{\odot}$,  
where the velocity dispersion of the members 
$\sigma_v$ is $365\pm34$ km s$^{-1}$ and the spatial extent $R$ is $\approx 180$ kpc. 
Similary, the LABs can be bound to objects with mass $\ga 10^{10.5}~M_{\odot}$, 
although it is lower than the total stellar masses 
of the LABs of $M_{\ast}=0.4\sim2.0\times10^{11}~M_{\odot}$. 

\section{Discussion}
\subsection{Mass constraints of the groups}

The observed velocity distributions of the group members 
show only a part of their dynamical properties. 
First, the observed velocities and spatial distributions 
of the galaxies are the projected ones. 
Second, the groups probably represent 
the properties of only the central parts of their host halos.

To evaluate the physical properties of the groups, 
we compare our samples with the galaxy groups selected from 
the galaxy formation models based on the Millennium simulation 
(\citealt{2005Natur.435..629S}). 
We assume that the observed groups represent the central dense parts of virialised halos. 
We select the comparison groups which contain the same number
of galaxies with the stellar mass larger than the completeness limit as our sample groups
from the groups hosted within the collapsed halos at $z\sim3$ (at the $z=2.86$ and $z=3.06$ snapshots ) 
in a comoving volume of $1.25\times10^8~h^{-3}$ Mpc$^3$ of the full Millennium simulation. 
The comparison Millennium groups are also within the same angular sizes as our sample groups. 
For example, a group containing five galaxies with $M_{\ast}\geq 4\times 10^{10}~M_{\odot}$
within 120 kpc from their centre is selected as the comparison group for the AzTEC14 group. 
We use the galaxy catalogues based on the galaxy formation models
of \citet{2007MNRAS.375....2D} and \citet{2011MNRAS.413..101G}. 
In the latter model, the efficiency of galaxy formation, especially in low mass halos 
decreases to fit the observed values. 
The Chabrier IMF is adopted in both the galaxy formation models. 

Fig. \ref{fig:space-velocity-2} shows the comparisons 
of the spatial and velocity distributions of our samples 
and the comparison Millennium groups at $z\sim3$ selected from \citet{2011MNRAS.413..101G}. 
Light grey points show the physical spatial and velocity separations 
of the members of the comparison Millennium groups,
and dark grey points with the same symbols show those seen from one direction. 
We show the mean physical spatial separations expected from their angular separations,  
namely the angular separations multiplied by $\pi/2$, 
and the mean velocity offsets seen from one direction 
expected from their physical velocity separations, 
i.e. the physical velocity separations multiplied by $4/\pi^2$.
Note that for the AzTEC14 group, only one comparison group 
is found in each $z\sim3$ snapshot of the full Millennium volume 
and such groups are only found at the centres of collapsed massive halos. 
It is because, as we describe later, 
massive halos which can host AzTEC14-like groups are very rare at $z\sim3$. 
The velocity distributions of the AzTEC14 group 
and LAB30 closely resemble to those of the projected comparison Millennium groups.   
On the other hand, the velocities of the galaxies associated with LAB01 
are on average lower than those in the comparison Millennium groups. 
This suggests that LAB01 is actually hosted in a lower mass halo or 
associated with the galaxies which are dynamically bound 
but just falling into the potential of the central high density region. 

\begin{figure}
\begin{center}
\includegraphics[width=82mm]{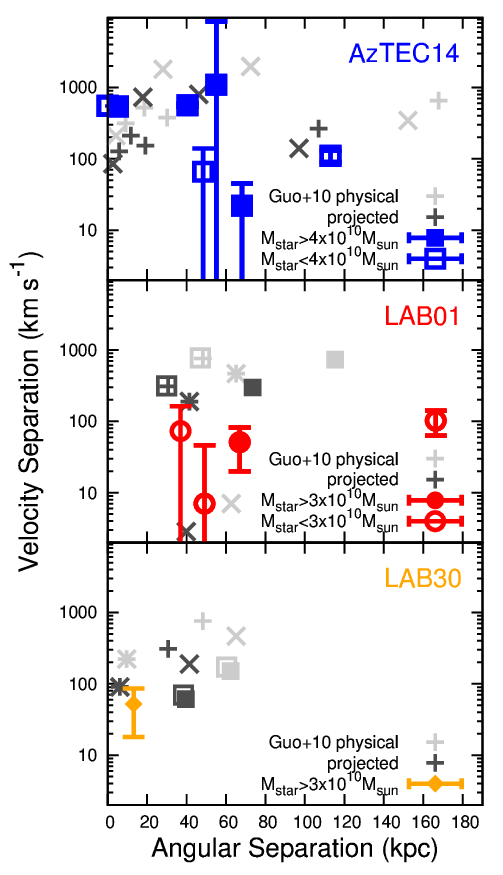}
\caption{
 The comparison of the velocity and spatial distributions of our samples
 and the galaxy groups selected from the galaxy formation models based on the Millennium simulation. 
Top panel: Blue filled and blank squares show the members of the AzTEC14 group 
with $M_{\ast}\geq 4\times10^{10}~M_{\odot}$ and  $M_{\ast}<4\times10^{10}~M_{\odot}$, respectively. 
Light grey cross and x-mark points show the physical separations and velocity distributions 
of the galaxies with $M_{\ast}\geq 4\times10^{10}~M_{\odot}$ 
in the comparison Millennium groups selected
from the $z=2.86$ and $z=3.06$ snapshots of \citet{2011MNRAS.413..101G}, respectively. 
Dark grey points shown with the same symbols as the light grey points 
show the physical separations and velocity distributions seen from one direction.
Middle panel:  Similar to the top panel but 
red filled and blank circles show the counterparts of LAB01 
with $M_{\ast}\geq 3\times10^{10}~M_{\odot}$ and  $M_{\ast}<3\times10^{10}~M_{\odot}$, respectively. 
Light grey and dark grey points are similar to those in the top panel 
but for the galaxies with $M_{\ast}\geq 3\times10^{10}~M_{\odot}$ 
in the comparison Millennium groups selected from the $z=3.06$ snapshot. 
Bottom panel: Similar to the middle panel but for LAB30 shown with an orange diamond. 
} 
\label{fig:space-velocity-2}
\end{center} 
\end{figure}

Fig. \ref{fig:smdist} shows the stellar mass distributions 
of the AzTEC14 group, LAB01 and their comparison groups 
in \citet{2007MNRAS.375....2D} and \citet{2011MNRAS.413..101G}. 
The observed stellar mass distributions agree well with both the Millennium models
at above the completeness limit.  
Indeed, the comparison of our groups and the Millennium groups 
is reasonable at above this limit, although there can be further differences at low mass galaxies.  
As e.g., \citet{2012ApJ...752L..19Q} reported,  
the numbers of satellite galaxies around the local massive galaxies 
are smaller than those predicted by the galaxy formation models based on the Millennium simulation. 

We now consider the masses of the halos which host such multiple massive galaxies.
At each $z\sim3$ snapshot of the Millennium simulation, 
only one massive halo hosts the comparison group of the AzTEC14 group;
The halos with the virial mass $M_{\rm vir} \approx 10^{13.9}(10^{14.0})~M_{\odot}$ 
and $10^{13.9}(10^{13.6})~M_{\odot}$ at the $z=3.06~($ or at the $z=2.86)$ snapshot 
respectively host the comparison groups in the \citet{2007MNRAS.375....2D} and \citet{2011MNRAS.413..101G} models. 
In the \citet{2007MNRAS.375....2D} model, the comparison 
Millennium group of the AzTEC14 group at $z=2.86$ is the descendant
of the comparison group at $z=3.06$. 
In the \citet{2011MNRAS.413..101G} model,
another massive halo happens to host the comparison group of the AzTEC14 group at $z=2.86$ 
while the number of the galaxies in the descendent 
of the comparison group at $z=3.06$ decreases by mergers.
Since the comparison groups of the AzTEC14 group are very rare at $z\sim3$, 
we also see the halo mass distribution of such groups 
by rescaling the galaxy groups at low redshift to be at $z=3$. 
We select the galaxy groups which contain the same numbers of galaxies 
within the same angular sizes as the AzTEC14 group in the comoving scale
from the $z=0-0.5$ snapshots of the Millennium simulation, 
assuming that the virial masses of the halos are preserved 
but their sizes evolve as $\propto (1+z)^{-1}$. 
The local comparison groups of the AzTEC14 group are also hosted in very massive halos 
while their host halo masses range from $M_{\rm vir}=10^{13.7}$ to $10^{14.0}~M_{\odot}$ 
with median $10^{13.9}~M_{\odot}$ for the \citet{2007MNRAS.375....2D} model
and from $M_{\rm vir}=10^{13.6}$ to $10^{14.2}~M_{\odot}$ 
with median $10^{13.9}~M_{\odot}$ for the \citet{2011MNRAS.413..101G} model. 

Conversely, LAB01 and LAB30 host two galaxies with $M_{\ast}\geq 3\times10^{10}~M_{\odot}$ 
within 80 kpc and 60 kpc, respectively.
The masses of the Millennium halos hosting the comparison groups at $z=3.06$
range from $M_{\rm vir}=10^{12.2}$ to $10^{14.0}~M_{\odot}$ 
with median $10^{13.2}~M_{\odot}$ 
for the \citet{2007MNRAS.375....2D} model and from $M_{\rm vir}=10^{12.4}$ to $10^{14.1}~M_{\odot}$ 
with median $10^{13.2}~M_{\odot}$ for the \citet{2011MNRAS.413..101G} model. 
These are similar to those of the host halos 
of the comparison groups selected at low redshift with rescaling. 

From these results, we argue that the groups found 
in the SSA22 protocluster match well with the groups 
hosted in massive halos at $z\sim3$ in the Millennium simulation.
By tracing the evolution of a comparison group for the AzTEC14 group from $z\sim3$, 
most of its members merge into one massive galaxy 
at the centre of its host halo while the host halo 
becomes 10 times massive in the local Universe. 
This predicts that the AzTEC14 group will evolve into the BCG 
of one of the most massive clusters in the local Universe, 
whereas the SSA22 protocluster itself is thought to be a progenitor 
of one of the most massive clusters (\citealt{2012AJ....143...79Y}; K15). 
The comparison groups of LAB01 and LAB30 at $z\sim3$ 
also mostly merge into massive galaxies at $z=0$. 

\begin{figure*}
\begin{center}
\includegraphics[width=150mm]{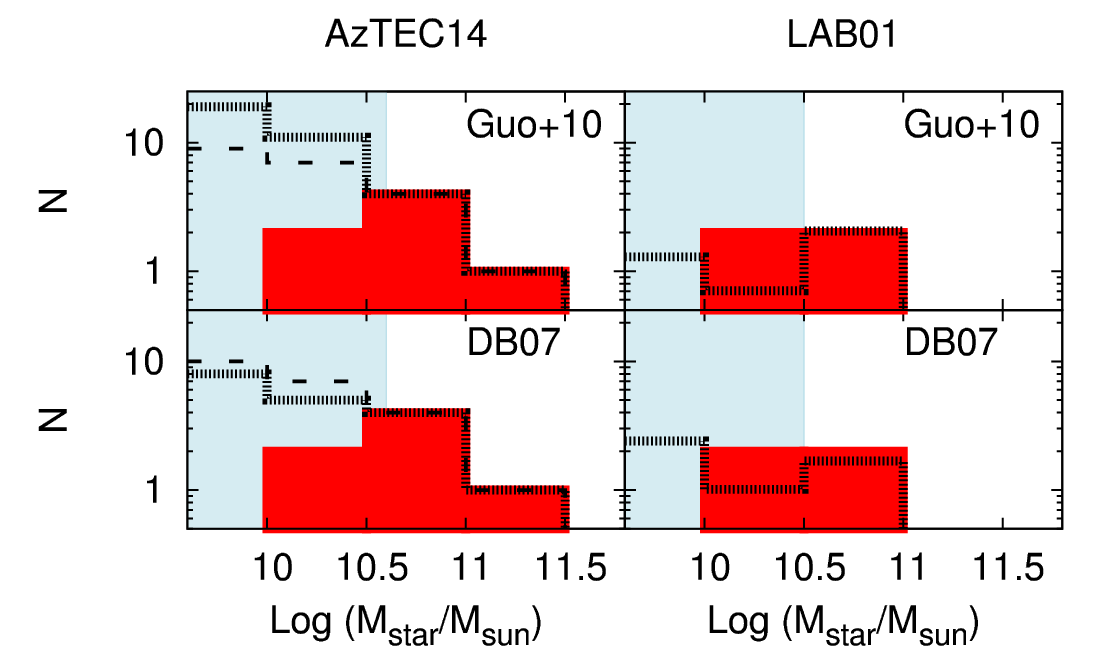}
\caption{
Left top panel: Red filled histogram shows the stellar mass distribution of the AzTEC14 group.  
The histograms drawn with black dashed and dotted lines are those 
of the comparison Millennium groups selected from the snapshots 
at $z=2.86$ and $z=3.06$ of the \citet{2011MNRAS.413..101G} model, respectively. 
Grey shaded region indicates the stellar mass range below the completeness limit 
estimated from the observed stellar populations of the counterparts of the AzTEC14 group. 
Left bottom panel: Similar to the left top panel but 
the comparison Millennium groups are selected from the \citet{2007MNRAS.375....2D} model. 
Right top panel: Similar to the left top panel but the red histogram shows the stellar mass distribution of LAB01.
The histogram drawn with a black dotted line shows the averaged stellar mass distribution 
of the comparison Millennium groups selected 
from the $z=3.06$ snapshot of the \citet{2011MNRAS.413..101G} model. 
Right bottom panel: Similar to the right top panel 
but the comparison Millennium groups are selected from the \citet{2007MNRAS.375....2D} model. 
} 
\label{fig:smdist}
\end{center} 
\end{figure*}

\subsection{Assembly histories of the most massive galaxies}

As we discussed above, we are very likely to see multiple merging phases 
in the formation history of massive galaxies. 
In this subsection, we discuss their assembly histories.  

As mentioned earlier, the most plausible descendant of the AzTEC14 group
is a BCG with $M_{\ast}\sim 10^{12}~M_{\odot}$ hosted within one of the most massive clusters. 
Stellar mass of $10^{12}~M_{\odot}$ is a sort of maximum while 
the stellar mass of a BCG depends on the mass 
of its host cluster (e.g., \citealt{2012MNRAS.422.2213S}). 
Over half the stellar mass of its descendant 
is present at $z\sim3$, when the total stellar mass of the AzTEC14 group 
is already $M_{\ast}=5.8^{+5.1}_{-2.0}\times10^{11}~M_{\odot}$.  
Given that the SEDs of Az14-K15a and Az14-K15c are well characterised 
as those of the galaxies with burst-like star formation histories and ages of $0.3-2$ Gyr (K15),  
a third of the stellar mass of the descendent of the AzTEC14 group may have been formed by $z\sim3.6$. 
On the other hand, the total stellar masses of the LABs 
range from $M_{\ast}=0.4-2.0\times10^{11}~M_{\odot}$, 
which are already comparable with over half the stellar masses of massive ellipticals in local clusters. 
Our results support the early formation 
of the stars of massive ellipticals in the form of multiple progenitors,  
as predicted by cosmological numerical simulations 
(e.g., \citealt{2003ApJ...590..619M}; \citealt{2007ApJ...658..710N}; 
\citealt{2007MNRAS.375....2D}; \citealt{2010ApJ...725.2312O}). 

The presence of massive quiescent galaxies implicates that 
some quenching mechanisms must have already been invoked for at least a part of the groups. 
Large velocity offsets of Ly$\alpha$ from systemic redshifts in 
some members of LABs indicate the strong outflow. 
In addition, AGNs are found in the massive quiescent 
counterparts of LAB12 and the AzTEC14 group, 
although their AGN activities are only slightly higher than 
those of massive ellipticals in the local Universe (K15). 
Moreover, since they are likely to be hosted 
in massive halos with $M_{\rm halo}\ga 10^{13}~M_{\odot}$, 
gravitational heating can globally suppress
their star formation activities (e.g., \citealt{2008MNRAS.383..119D}). 

It is also interesting to discuss how SMBHs have grown in the groups. 
In our sample, LAB12-K15 and AzTEC14-K15a are detected in X-ray with {\it Chandra} \citep{2009MNRAS.400..299L}. 
Moreover, the large line widths of some counterparts suggest the presence of AGNs 
while the instrumental resolution corrected velocity dispersions  
estimated from [O{\footnotesize III}] $\lambda$5007 
of LAB01-K15a, LAB12-K15 and AzTEC14-K15a are 
$111\pm43$ km s$^{-1}$, $139\pm19$ km s$^{-1}$, 
and $225\pm11$ km s$^{-1}$ (shorter) and $102\pm17$ km s$^{-1}$ (longer), respectively, 
which are larger than those of the typical galaxies at $z=2-3$ 
with similar stellar mass \citep{2006ApJ...646..107E}.
Interestingly, the most massive member of the AzTEC14 group, 
AzTEC14-K15c shows no significant detection in X-ray. 
X-ray luminosity is given as 
$L_X =2\times10^{43}~(M_{\rm BH}/10^8~M_{\odot})(\epsilon_X/0.01)(R_{\rm edd}/0.1)$ erg s$^{-1}$ 
where $\epsilon_{X}$ is the bolometric fraction of the X-ray emission at $2-10$ keV 
and $R_{\rm edd}$ is the Eddington ratio \citep{2009ApJ...699.1354Y}.
Assuming $M_{\rm BH}/M_{\rm sph}\approx0.002$ of local value, $\epsilon_{X}=0.01$ and $R_{\rm edd}=0.1$, 
Az14-K15c can have $L_{X} \sim 1 \times 10^{44}$ erg s$^{-1}$,   
which is larger than the detection limit of the existing {\it Chandra} data 
$\sim10^{42.6}$ erg s$^{-1}$ at $2-8$ keV, 
which suggests that its $\epsilon_{X}$, $R_{\rm edd}$ and/or $M_{\rm BH}/M_{\rm sph}$ ratio are several times lower than the above values. 
Since AzTEC14-K15c is well characterized as a massive quiescent galaxy, its accretion rate may be low. 
Therefore subsequent mergers of BHs and/or accretion of matter 
induced by galaxy mergers may be required to grow its SMBH. 
Such mergers accompanied with SMBHs are of particular importance 
that they can heat to reduce the central stellar mass 
by energy release during the orbital decays of binary SMBHs 
and reproduce the observed core structures of massive ellipticals 
(e.g., \citealt{1980Natur.287..307B, 2003ApJ...582..559V}). 

\subsection{LABs and SMGs} 
The origin of LABs is a long argued mystery. 
Several scenarios have been proposed as the origin of their extended Ly$\alpha$ halos, 
like hidden QSOs (e.g., \citealt{2001ApJ...556...87H}), obscured starburst with strong super-winds 
induced by death of massive stars (e.g., \citealt{2000ApJ...532L..13T}) 
and release of gravitational energy from cold streams (e.g., \citealt{2008MNRAS.383..119D}).
Our results verify the argument of \citet{2012ApJ...750..116U} 
that the substantial fraction of the LABs in the SSA22 protocluster 
are associated with massive stellar components. 
Interestingly, the locations of the stellar components 
do not always match with the surface brightness peaks of Ly$\alpha$. 
This suggests that some of the Ly$\alpha$ is possibly emitted from 
and/or scattered by neutral hydrogen gas,  
and/or ionized by leaked emission from obscured sources.

It is also interesting to compare the properties 
of the galaxy groups associating with SMGs and LABs. 
Both objects are plausible progenitors of massive galaxies in multiple merging phases 
in the SSA22 protocluster. 
Some LABs in the SSA22 field are significantly detected 
in the SCUBA 850 $\mu$m \citep{2005MNRAS.363.1398G} 
and  the AzTEC/ASTE 1.1 mm \citep{2013MNRAS.430.2768T}
with the yielded mean 1$\sigma$ noise limit of $S_{850\rm~\mu m}=1.5$ mJy 
and an rms noise level of $S_{\rm 1.1~mm}=0.7-1$ mJy beam$^{-1}$, respectively. 
On the other hand, most of the LABs are not likely to be bright sub-mm sources; 
It was reported by \citet{2013MNRAS.430.2768T} that 
most of the LABs in the SSA22 field are not significantly detected 
in the AzTEC/ASTE 1.1 mm survey and their stacked 1.1 mm 
flux density is $S_{1.1\rm~mm}<0.40$ mJy (3$\sigma$).  

Both the AzTEC14 group and LAB01
have multiple massive stellar components and large spatial extents  
but the former shows no significant Ly$\alpha$ emission. 
LAB01 is associated with an 850 mm source 
with $S_{\rm 850~\mu m}=(4.6\pm1.1)$ mJy \citep{2014ApJ...793...22G} 
and the 1.1 mm flux density of LAB01 is 1.97$\pm0.74$ mJy 
(S/N=2.7, \citealt{2013MNRAS.430.2768T}) 
while that of SSA22-AzTEC14 is $S_{\rm 1.1~mm}=(5.0\pm0.7)$ mJy \citep{2014MNRAS.440.3462U}. 
There are several possible scenarios inducing the Ly$\alpha$ deficiency in the group scale. 
There can be further supply of dust among the AzTEC14 group  
and LAB12, which has a sparse Ly$\alpha$ nebula, 
since they host well characterised massive quiescent galaxies as old as  
the age at which the dust ejections from AGB stars become effective \citep{2009Sci...323..353S}. 
The large velocity dispersion of 
the AzTEC14 group implies that the dust and gas geometry in the group has been mixed 
much more by interactions and mergers of galaxies than those in LAB01.  
Moreover, gravitational heating of a halo can overcome the cooling by filamentary 
cold streams \citep{2008MNRAS.383..119D} in the AzTEC14 group. 
Although it may not be the general case since this group is a very rare massive object at $z\sim3$. 
Further studies with ALMA will help us to solve 
the different nature of the galaxy groups coexisting in such a dense environment.

\section{CONCLUSIONS}

We discovered a very dense physically associated group 
of galaxies as the counterparts of a SMG in the SSA22 
protocluster at $z=3.09$ called the AzTEC14 group.   
By comparing with the galaxy formation models based on the Millennium simulation, 
we found that the AzTEC14 group shows similar properties as those of the Millennium groups 
hosted in halos with $M_{\rm vir}=10^{13.4}-10^{14.0}~M_{\odot}$ at $z\sim3$.
Most of the members of such a Millennium group at $z\sim3$ merge
into one central massive galaxy 
at the centre of a halo that is 10 times massive in the current Universe. 
We also confirmed two or more counterparts of the four LABs.  
Similarly, they are comparable with the Millennium groups 
hosted in halos with median mass $M_{\rm vir}\sim 10^{13.2}~M_{\odot}$. 
The total stellar masses of the AzTEC14 group and the LABs 
are already larger than about half the stellar masses of 
BCGs or massive ellipticals in the local Universe, 
suggesting the early formation of stars in each group. 
Our results strongly support the hierarchical formation 
of massive galaxies predicted in the $\Lambda$CDM cosmology. 

\section*{Acknowledgments}

This study is based on data collected at Subaru Telescope, 
which is operated by the National Astronomical Observatory of Japan. 
We would like to thank the Subaru Telescope staff 
for many help and support for the observations. 
Our studies owe a lot deal to the archival Subaru 
Suprime-Cam (\citealt{2004AJ....128..569M}), {\it Spitzer} IRAC \& MIPS data taken in
\citet{2009ApJ...692.1561W}, {\it Chandra} data taken in
\citet{2009ApJ...691..687L}. 
We also thank to AzTEC/ASTE observers of the SSA22 field 
providing the updated source catalog.  
This work was supported by Global COE Program "Weaving
Science Web beyond Particle-Matter Hierarchy", MEXT, Japan. 
YM acknowledges support from JSPS KAKENHI Grant Number 20647268. 
This work was partially supported by JSPS Grants-in-Aid for Scientific  Research No.26400217.

\appendix
\section{The best-fit SED models of the spectroscopically confirmed counterparts of the AzTEC14 group and the LABs}
\begin{figure*}
\begin{center}
\includegraphics[width=175mm]{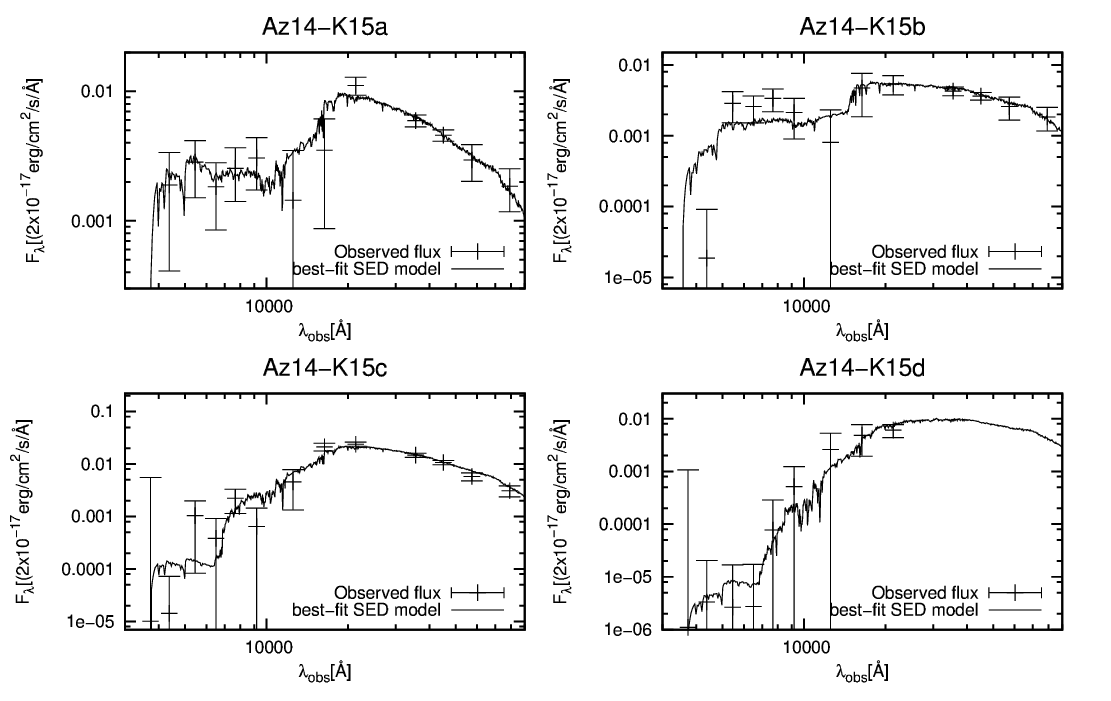}
\includegraphics[width=175mm]{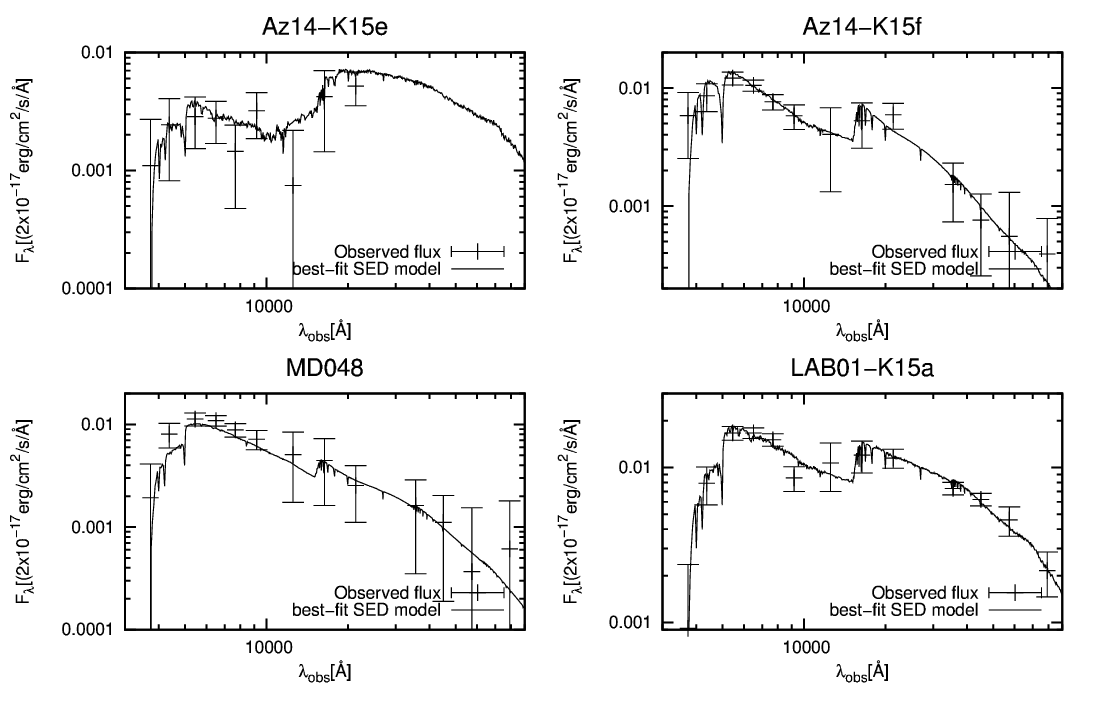}
\caption{
The observed spectra and their best-fit SEDs of the counterparts 
of the AzTEC14 group and the LABs.
Cross points show the flux values observed 
in the $u^{\star}BVRi'z'JHK$, 3.6, 4.5, 5.8 and 8.0 $\mu$m-bands. 
The solid lines show their best-fit SED models. 
At the missing data points, flux values cannot be constrained 
since the sources are deblended with the neighboring sources. 
The $J$-band flux values get out of the best-fit SED models largely for some objects. 
There is no obvious deblending but since our $J$-band images are very deep, 
it can happen that some other faint unresolved components detected in $J$-band only,  
like blue faint galaxies and [O{\scriptsize II}] $\lambda$3727 emitters, 
are deblended with the central objects. 
} 
\label{fig:sedsll}
\end{center} 
\end{figure*}
\addtocounter{figure}{-1}
\begin{figure*}
\begin{center}
\includegraphics[width=175mm]{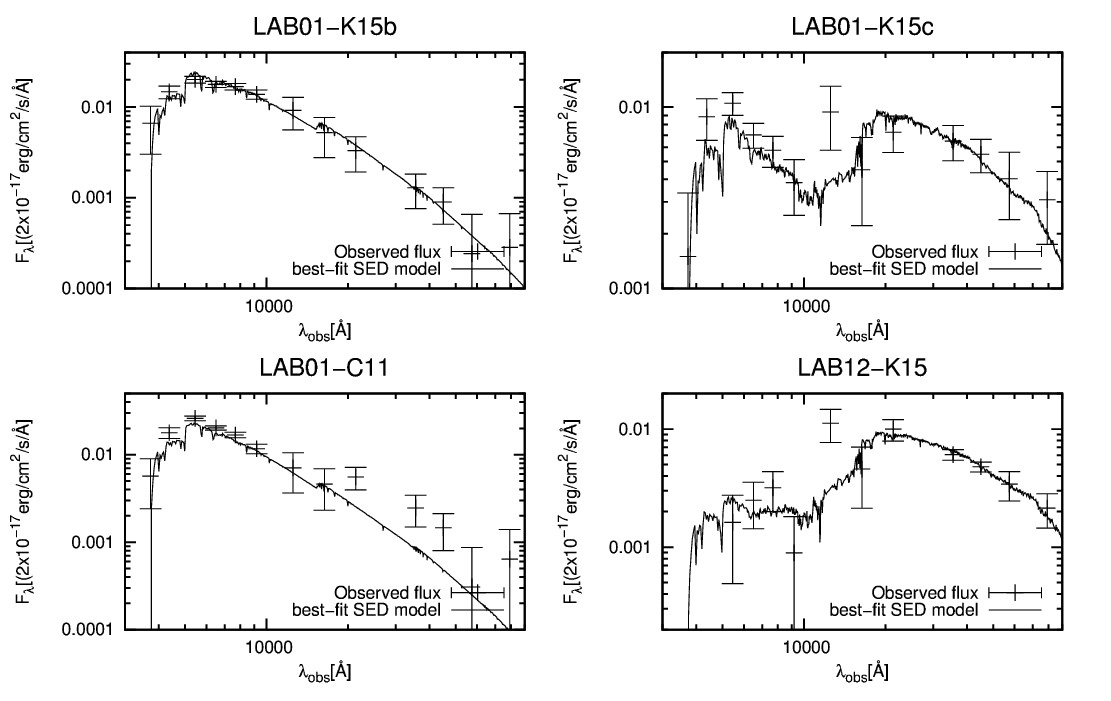}
\includegraphics[width=175mm]{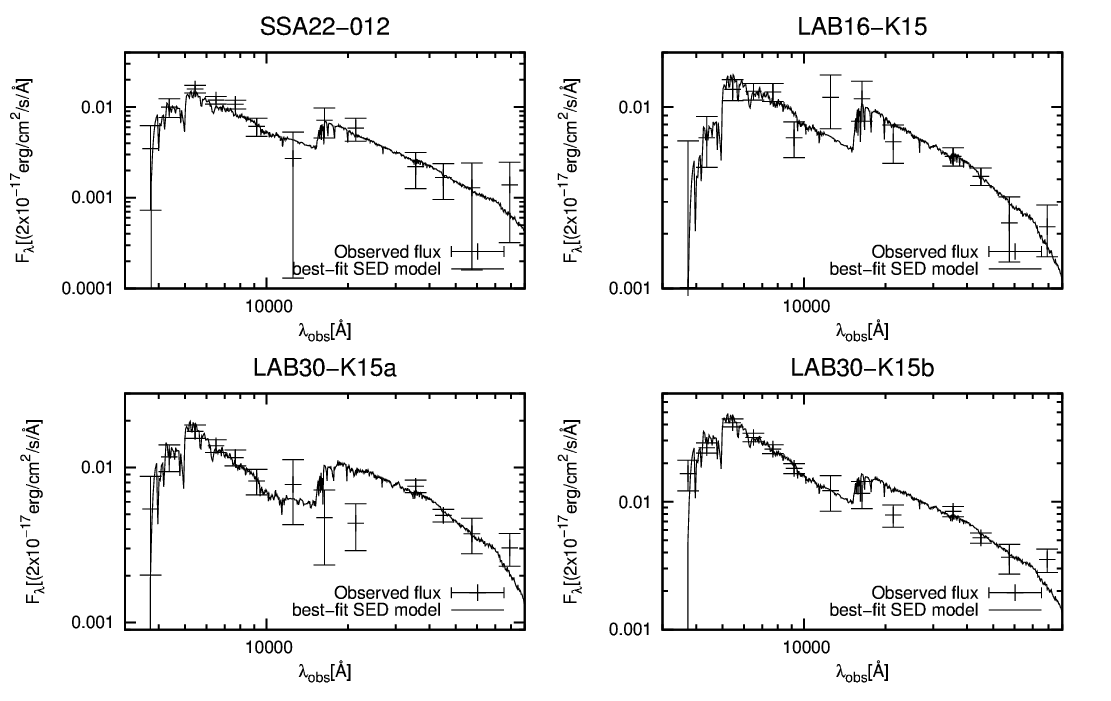}
\caption{
{\it Continued.} 
} 
\end{center} 
\end{figure*}

\label{lastpage}


\begin{thebibliography}{99}
\bibitem[\protect\citeauthoryear{Begelman, Blandford,
\& Rees}{1980}]{1980Natur.287..307B} Begelman M.~C., Blandford R.~D., Rees M.~J., 1980, Natur, 287, 307

\bibitem[\protect\citeauthoryear{Bezanson et
al.}{2009}]{2009ApJ...697.1290B} Bezanson R., van Dokkum P.~G., Tal T.,
Marchesini D., Kriek M., Franx M., Coppi P., 2009, ApJ, 697, 1290

\bibitem[\protect\citeauthoryear{Bruzual
\& Charlot}{2003}]{2003MNRAS.344.1000B} Bruzual G., Charlot S., 2003, MNRAS, 344, 1000


\bibitem[\protect\citeauthoryear{Calzetti et
al.}{2000}]{2000ApJ...533..682C} Calzetti D., Armus L., Bohlin R.~C.,
Kinney A.~L., Koornneef J., Storchi-Bergmann T., 2000, ApJ, 533, 682


\bibitem[\protect\citeauthoryear{Chabrier}{2003}]{2003PASP..115..763C}
Chabrier G., 2003, PASP, 115, 763


\bibitem[\protect\citeauthoryear{Croton et al.}{2006}]{2006MNRAS.365...11C}
Croton D.~J., et al., 2006, MNRAS, 365, 11

\bibitem[\protect\citeauthoryear{Daddi et al.}{2005}]{2005ApJ...626..680D}
Daddi E., et al., 2005, ApJ, 626, 680

\bibitem[\protect\citeauthoryear{De Lucia et
al.}{2006}]{2006MNRAS.366..499D} De Lucia G., Springel V., White S.~D.~M.,
Croton D., Kauffmann G., 2006, MNRAS, 366, 499


\bibitem[\protect\citeauthoryear{De Lucia
\& Blaizot}{2007}]{2007MNRAS.375....2D} De Lucia G., Blaizot J., 2007, MNRAS, 375, 2


\bibitem[\protect\citeauthoryear{Dekel
\& Birnboim}{2008}]{2008MNRAS.383..119D} Dekel A., Birnboim Y., 2008, MNRAS, 383, 119


\bibitem[\protect\citeauthoryear{Erb et al.}{2006}]{2006ApJ...646..107E}
Erb D.~K., Steidel C.~C., Shapley A.~E., Pettini M., Reddy N.~A.,
Adelberger K.~L., 2006, ApJ, 646, 107


\bibitem[\protect\citeauthoryear{Erb et al.}{2014}]{2014ApJ...795...33E}
Erb D.~K., et al., 2014, ApJ, 795, 33


\bibitem[\protect\citeauthoryear{Franx et al.}{2003}]{2003ApJ...587L..79F}
Franx M., et al., 2003, ApJ, 587, L79


\bibitem[\protect\citeauthoryear{Geach et al.}{2005}]{2005MNRAS.363.1398G}
Geach J.~E., et al., 2005, MNRAS, 363, 1398

\bibitem[\protect\citeauthoryear{Geach et al.}{2014}]{2014ApJ...793...22G}
Geach J.~E., et al., 2014, ApJ, 793, 22


\bibitem[\protect\citeauthoryear{Guo et al.}{2011}]{2011MNRAS.413..101G}
Guo Q., et al., 2011, MNRAS, 413, 101


\bibitem[\protect\citeauthoryear{Haiman
\& Rees}{2001}]{2001ApJ...556...87H} Haiman Z., Rees M.~J., 2001, ApJ, 556, 87


\bibitem[\protect\citeauthoryear{Hayashino et
al.}{2004}]{2004AJ....128.2073H} Hayashino T., et al., 2004, AJ, 128, 2073


\bibitem[\protect\citeauthoryear{Hopkins et
al.}{2009}]{2009MNRAS.398..898H} Hopkins P.~F., Bundy K., Murray N.,
Quataert E., Lauer T.~R., Ma C.-P., 2009, MNRAS, 398, 898

\bibitem[\protect\citeauthoryear{Kennicutt}{1983}]{1983ApJ...272...54K}
Kennicutt R.~C., Jr., 1983, ApJ, 272, 54


\bibitem[\protect\citeauthoryear{Klypin, Trujillo-Gomez,
\& Primack}{2011}]{2011ApJ...740..102K} Klypin A.~A., Trujillo-Gomez S., Primack J., 2011, ApJ, 740, 102

\bibitem[\protect\citeauthoryear{Kubo et al.}{2013}]{2013ApJ...778..170K}
Kubo M., et al., 2013, ApJ, 778, 170


\bibitem[\protect\citeauthoryear{Kubo et al.}{2015}]{2015ApJ...799...38K}
Kubo M., Yamada T., Ichikawa T., Kajisawa M., Matsuda Y., Tanaka I., 2015,
ApJ, 799, 38


\bibitem[\protect\citeauthoryear{Laine et al.}{2003}]{2003AJ....125..478L}
Laine S., van der Marel R.~P., Lauer T.~R., Postman M., O'Dea C.~P., Owen
F.~N., 2003, AJ, 125, 478


\bibitem[\protect\citeauthoryear{Lauer}{1988}]{1988ApJ...325...49L} Lauer
T.~R., 1988, ApJ, 325, 49


\bibitem[\protect\citeauthoryear{Lehmer et al.}{2009}]{2009MNRAS.400..299L}
Lehmer B.~D., et al., 2009, MNRAS, 400, 299


\bibitem[\protect\citeauthoryear{Lehmer et al.}{2009}]{2009ApJ...691..687L}
Lehmer B.~D., et al., 2009, ApJ, 691, 687


\bibitem[\protect\citeauthoryear{MacKenzie et
al.}{2014}]{2014MNRAS.445..201M} MacKenzie T.~P., et al., 2014, MNRAS, 445,
201

\bibitem[\protect\citeauthoryear{Matsuda et
al.}{2004}]{2004AJ....128..569M} Matsuda Y., et al., 2004, AJ, 128, 569


\bibitem[\protect\citeauthoryear{Matsuda et
al.}{2005}]{2005ApJ...634L.125M} Matsuda Y., et al., 2005, ApJ, 634, L125


\bibitem[\protect\citeauthoryear{Matsuda et
al.}{2006}]{2006ApJ...640L.123M} Matsuda Y., Yamada T., Hayashino T.,
Yamauchi R., Nakamura Y., 2006, ApJ, 640, L123

\bibitem[\protect\citeauthoryear{Matsuda et
al.}{2009}]{2009MNRAS.400L..66M} Matsuda Y., et al., 2009, MNRAS, 400, L66

\bibitem[\protect\citeauthoryear{Matsuda et
al.}{2011}]{2011MNRAS.410L..13M} Matsuda Y., et al., 2011, MNRAS, 410, L13

\bibitem[\protect\citeauthoryear{McLinden et
al.}{2013}]{2013ApJ...767...48M} McLinden E.~M., Malhotra S., Rhoads J.~E.,
Hibon P., Weijmans A.-M., Tilvi V., 2013, ApJ, 767, 48

\bibitem[\protect\citeauthoryear{Meza et al.}{2003}]{2003ApJ...590..619M}
Meza A., Navarro J.~F., Steinmetz M., Eke V.~R., 2003, ApJ, 590, 619

\bibitem[\protect\citeauthoryear{Naab, Khochfar,
\& Burkert}{2006}]{2006ApJ...636L..81N} Naab T., Khochfar S., Burkert A., 2006, ApJ, 636, L81

\bibitem[\protect\citeauthoryear{Naab et al.}{2007}]{2007ApJ...658..710N}
Naab T., Johansson P.~H., Ostriker J.~P., Efstathiou G., 2007, ApJ, 658,
710

\bibitem[\protect\citeauthoryear{Naab, Johansson,
\& Ostriker}{2009}]{2009ApJ...699L.178N} Naab T., Johansson P.~H., Ostriker J.~P., 2009, ApJ, 699, L178

\bibitem[\protect\citeauthoryear{Navarro, Frenk,
\& White}{1997}]{1997ApJ...490..493N} Navarro J.~F., Frenk C.~S., White S.~D.~M., 1997, ApJ, 490, 493

\bibitem[\protect\citeauthoryear{Oser et al.}{2010}]{2010ApJ...725.2312O}
Oser L., Ostriker J.~P., Naab T., Johansson P.~H., Burkert A., 2010, ApJ,
725, 2312


\bibitem[\protect\citeauthoryear{Osterbrock}{1989}]{1989agna.book.....O}
Osterbrock D.~E., 1989, agna.book,


\bibitem[\protect\citeauthoryear{Overzier et
al.}{2013}]{2013ApJ...771...89O} Overzier R.~A., Nesvadba N.~P.~H.,
Dijkstra M., Hatch N.~A., Lehnert M.~D., Villar-Mart{\'{\i}}n M., Wilman
R.~J., Zirm A.~W., 2013, ApJ, 771, 89


\bibitem[\protect\citeauthoryear{Prescott et
al.}{2012}]{2012ApJ...752...86P} Prescott M.~K.~M., et al., 2012, ApJ, 752,
86


\bibitem[\protect\citeauthoryear{Quilis
\& Trujillo}{2012}]{2012ApJ...752L..19Q} Quilis V., Trujillo I., 2012, ApJ, 752, L19


\bibitem[\protect\citeauthoryear{Salpeter}{1955}]{1955ApJ...121..161S}
Salpeter E.~E., 1955, ApJ, 121, 161

\bibitem[\protect\citeauthoryear{Seigar, Lynam,
\& Chorney}{2003}]{2003MNRAS.344..110S} Seigar M.~S., Lynam P.~D., Chorney N.~E., 2003, MNRAS, 344, 110


\bibitem[\protect\citeauthoryear{Sloan et al.}{2009}]{2009Sci...323..353S}
Sloan G.~C., et al., 2009, Sci, 323, 353


\bibitem[\protect\citeauthoryear{Springel et
al.}{2005}]{2005Natur.435..629S} Springel V., et al., 2005, Natur, 435, 629


\bibitem[\protect\citeauthoryear{Steidel et
al.}{1998}]{1998ApJ...492..428S} Steidel C.~C., Adelberger K.~L., Dickinson
M., Giavalisco M., Pettini M., Kellogg M., 1998, ApJ, 492, 428

\bibitem[\protect\citeauthoryear{Steidel et
al.}{2000}]{2000ApJ...532..170S} Steidel C.~C., Adelberger K.~L., Shapley
A.~E., Pettini M., Dickinson M., Giavalisco M., 2000, ApJ, 532, 170

\bibitem[\protect\citeauthoryear{Steidel et
al.}{2003}]{2003ApJ...592..728S} Steidel C.~C., Adelberger K.~L., Shapley
A.~E., Pettini M., Dickinson M., Giavalisco M., 2003, ApJ, 592, 728


\bibitem[\protect\citeauthoryear{Steidel et
al.}{2014}]{2014ApJ...795..165S} Steidel C.~C., et al., 2014, ApJ, 795, 165


\bibitem[\protect\citeauthoryear{Stott et al.}{2012}]{2012MNRAS.422.2213S}
Stott J.~P., et al., 2012, MNRAS, 422, 2213


\bibitem[\protect\citeauthoryear{Tacconi et
al.}{2006}]{2006ApJ...640..228T} Tacconi L.~J., et al., 2006, ApJ, 640, 228


\bibitem[\protect\citeauthoryear{Tacconi et
al.}{2008}]{2008ApJ...680..246T} Tacconi L.~J., et al., 2008, ApJ, 680, 246


\bibitem[\protect\citeauthoryear{Tamura et al.}{2009}]{2009Natur.459...61T}
Tamura Y., et al., 2009, Natur, 459, 61


\bibitem[\protect\citeauthoryear{Tamura et al.}{2013}]{2013MNRAS.430.2768T}
Tamura Y., et al., 2013, MNRAS, 430, 2768


\bibitem[\protect\citeauthoryear{Taniguchi
\& Shioya}{2000}]{2000ApJ...532L..13T} Taniguchi Y., Shioya Y., 2000, ApJ, 532, L13

\bibitem[\protect\citeauthoryear{Totani et al.}{2001}]{2001ApJ...558L..87T}
Totani T., Yoshii Y., Iwamuro F., Maihara T., Motohara K., 2001, ApJ, 558,
L87

\bibitem[\protect\citeauthoryear{Trujillo et
al.}{2007}]{2007MNRAS.382..109T} Trujillo I., Conselice C.~J., Bundy K.,
Cooper M.~C., Eisenhardt P., Ellis R.~S., 2007, MNRAS, 382, 109


\bibitem[\protect\citeauthoryear{Uchimoto et
al.}{2008}]{2008PASJ...60..683U} Uchimoto Y.~K., et al., 2008, PASJ, 60,
683


\bibitem[\protect\citeauthoryear{Uchimoto et
al.}{2012}]{2012ApJ...750..116U} Uchimoto Y.~K., et al., 2012, ApJ, 750,
116


\bibitem[\protect\citeauthoryear{Umehata et
al.}{2014}]{2014MNRAS.440.3462U} Umehata H., et al., 2014, MNRAS, 440, 3462


\bibitem[\protect\citeauthoryear{van der Wel et
al.}{2014}]{2014ApJ...788...28V} van der Wel A., et al., 2014, ApJ, 788, 28

\bibitem[\protect\citeauthoryear{van Dokkum et
al.}{2008}]{2008ApJ...677L...5V} van Dokkum P.~G., et al., 2008, ApJ, 677,
L5

\bibitem[\protect\citeauthoryear{Volonteri, Haardt,
\& Madau}{2003}]{2003ApJ...582..559V} Volonteri M., Haardt F., Madau P., 2003, ApJ, 582, 559


\bibitem[\protect\citeauthoryear{Viero et al.}{2012}]{2012MNRAS.421.2161V}
Viero M.~P., et al., 2012, MNRAS, 421, 2161


\bibitem[\protect\citeauthoryear{Webb et al.}{2009}]{2009ApJ...692.1561W}
Webb T.~M.~A., Yamada T., Huang J.-S., Ashby M.~L.~N., Matsuda Y., Egami
E., Gonzalez M., Hayashino T., 2009, ApJ, 692, 1561


\bibitem[\protect\citeauthoryear{Weijmans et
al.}{2010}]{2010MNRAS.402.2245W} Weijmans A.-M., Bower R.~G., Geach J.~E.,
Swinbank A.~M., Wilman R.~J., de Zeeuw P.~T., Morris S.~L., 2010, MNRAS,
402, 2245

\bibitem[\protect\citeauthoryear{White
\& Rees}{1978}]{1978MNRAS.183..341W} White S.~D.~M., Rees M.~J., 1978, MNRAS, 183, 341


\bibitem[\protect\citeauthoryear{Yamada et al.}{2002}]{2002ApJ...577L..89Y}
Yamada T., Koyama Y., Nakata F., Kajisawa M., Tanaka I., Kodama T., Okamura
S., De Propris R., 2002, ApJ, 577, L89

\bibitem[\protect\citeauthoryear{Yamada et al.}{2009}]{2009ApJ...699.1354Y}
Yamada T., et al., 2009, ApJ, 699, 1354


\bibitem[\protect\citeauthoryear{Yamada et al.}{2012}]{2012AJ....143...79Y}
Yamada T., Nakamura Y., Matsuda Y., Hayashino T., Yamauchi R., Morimoto N.,
Kousai K., Umemura M., 2012, AJ, 143, 79


\bibitem[\protect\citeauthoryear{Yang et al.}{2009}]{2009ApJ...693.1579Y}
Yang Y., Zabludoff A., Tremonti C., Eisenstein D., Dav{\'e} R., 2009, ApJ,
693, 1579

\bibitem[\protect\citeauthoryear{Yang et al.}{2012}]{2012ApJ...744..178Y}
Yang Y., et al., 2012, ApJ, 744, 178


\bibitem[\protect\citeauthoryear{Yang et al.}{2014}]{2014ApJ...793..114Y}
Yang Y., Zabludoff A., Jahnke K., Dav{\'e} R., 2014, ApJ, 793, 114



\end{thebibliography}
\end{document}